\documentclass[prd,aps,twocolumn]{revtex4}
\usepackage{latexsym}
\usepackage{amssymb}
\begin{document}
\newcommand{\beq}{\begin{equation}}
\newcommand{\eeq}{\end{equation}}
\newcommand{\beqa}{\begin{eqnarray}}
\newcommand{\eeqa}{\end{eqnarray}}
\newcommand{\Prd}{Phys. Rev D}
\newcommand{\Prl}{Phys. Rev. Lett.}
\newcommand{\Plb}{Phys. Lett. B}
\newcommand{\Cqg}{Class. Quantum Grav.}
\newcommand{\Np}{Nuc. Phys.}
\newcommand{\Ap}{Ann. Phys.}
\newcommand{\Ptp}{Prog. Theor. Phys.}
\newcommand{\half}{\frac 1 2\;}
\newcommand{\dg}{\digamma}

\title{Reply to {\em Acausality of Massive Charged Spin 2 Fields}}
\author{M. Novello, S. E. Perez Bergliaffa, and R. P. Neves}
\affiliation{ Centro Brasileiro de Pesquisas F\'{\i}sicas\\Rua
Dr.\ Xavier Sigaud 150, Urca 22290-180\\ Rio de Janeiro,
Brazil}
\date{\today}

\begin{abstract}
We show that, contrary to the claim in hep-th/0304050, the
propagation of a spin-2 field in an electromagnetic background is
{\em causal}.
\end{abstract}


\maketitle

 \renewcommand{\thefootnote}{\arabic{footnote}}

Recently \cite{nosso} we showed that the propagation of a spin 2
field in an electromagnetic background is causal. This assertion
is against results obtained previously (see \cite{d2} and
references therein). For the proof we used
used the first-order representation for
the equations of a spin 2 field, which is based on a three-index
tensor $F_{\mu\nu\alpha}$. The first
order formalism is equivalent to the more usual
second order formalism, as can be shown from
the results in \cite{nosso}. Another ingredient for
the proof was Hadamard's method
of
discontinuities, in which
$$
[F_{\mu\nu\alpha ;\lambda}] = k_\lambda f_{\mu\nu\alpha}.
$$
on the surface of discontinuity. The authors of
\cite{com}
claim that
\begin{itemize}

\item from the equations $k_\mu f^\mu =0$ and $f_\mu f^\mu =0$,
where $f_\mu = f_{\mu\nu\alpha} \eta^{\nu\alpha}$, it cannot be
concluded that $k_\mu$ is lightlike, and

\item that the masless gauge invariance that appears in the
eikonal limit, used in \cite{nosso} to {\em gauge away
acausalities} cannot be used in that way, because the
underlying massive theory is not gauge invariant.

\end{itemize}

It was shown in \cite{com} that the first claim is correct,
but no proof or support was given for the second claim. We will give here
a proof of the causal behaviour based solely, as we shall see,
in
the symmetry
present only in the equations for
the discontinuities.

As a result of taking the discontinuity of the equations of motion
for a spin 2 field in an electromagnetic background, we got in
\cite{nosso} the following equations:
\beq
f_{\mu\nu\alpha}k^\mu =0,
\eeq
\beq
f_{\mu\nu\alpha}k^\alpha =0,
\eeq
\beq
f_{\alpha\beta\lambda}f^\lambda k^2 = 0,
\eeq
\beq
A_{\alpha\beta ,\mu}f^{\alpha\beta\mu} = 0,
\eeq
\beq
\frac 3 2 \;ie\; A^{\alpha\beta}f_{\alpha\beta\mu}-m^2f_\mu = 0,
\eeq
\begin{eqnarray}
{{f_{\alpha\beta}}^{\lambda}}{}k_\mu + {{f_{\beta\mu}}^{\lambda}}%
{}k_\alpha + {{f_{\mu\alpha}}^{\lambda}}{}k_{\beta} -\frac{1}{2}\;
\delta^{\lambda}_{\;\alpha} f_{[\mu}k_{\beta ]}
   &   \nonumber \\
 - \frac{1}{2} \;\delta^{\lambda}_{\;\mu} f_{[\beta} k_{\alpha ]}
  - \frac{1}{2} \;\delta^{\lambda}_{\;\beta} f_{[\alpha}k_{,\mu ]}  =
  0&
 \label{star22}
\end{eqnarray}

Notice that the dependence of all these equations on
the polarization $\epsilon_{\mu\nu}$ is only through the tensor
$f_{\alpha\beta\mu}$, defined by
 \beq
 2f_{\alpha\mu\nu} =
\epsilon_{\nu\alpha}k_{\mu}- \epsilon_{\nu\mu}k_{\alpha} +
f_\alpha \eta_{\mu\nu} - f_\mu \eta_{\alpha\nu}, \label{discf}
\eeq
and its trace. It is patently evident that equations (1)-(6)
are {\em invariant} under the transformation
\beq
\tilde \epsilon_{\mu\nu} = \epsilon_{\mu\nu} + \Lambda k_\mu k_\nu,
\label{trafo}
\eeq
where $\Lambda $ is an arbitrary function of
the coordinates. Note that this transformation is
{\em not} a symmetry of the massless theory in the presence
of the electromagnetic field, as can be shown by explicit calculation.
It is instead a symmetry of the equations for the discontinuity of the derivative
of the field $F_{\mu\nu\alpha}$. It follows from this symmetry
that the modulus of $X_\mu \equiv \epsilon_{\mu\nu}k^\nu$
transforms as \cite{nosso}
\beq
\tilde X^2 = X^2 + \Lambda (k^2)^2( 2\epsilon + \Lambda ).
\eeq
Depending on $\Lambda$, the vector $X_\mu$
(and any other projection of the polarization tensor)
may be timelike, null, or spacelike. Only for the case
$k^2=0$ this unacceptable dependence of an observable quantity with a gauge choice
disappears . We conclude then that the propagation
of a spin-2 field in an electromagnetic background is
{\em causal}.

\end{document}